\documentclass[twocolumn,aps,amsmath,nofootinbib]{revtex4}
\usepackage{graphicx}
\usepackage{amssymb}
\usepackage{color}
\definecolor{joerg}{rgb}{1.0,0.0,0.0}

\newcommand{\eq}{\begin{equation}}
\newcommand{\feq}{\end{equation}}

\allowdisplaybreaks[1]

\def\calo{\mathcal{O}}

\newcommand{\be}{\begin{equation}}
\newcommand{\ee}{\end{equation}}

\makeatother

\begin{document}
\title{Ultraviolet properties of $f(R)$--Gravity}
\author{Alessandro Codello}
\email{a.codello@gmail.com}
\affiliation{SISSA, via Beirut 4, I-34014 Trieste, Italy, and INFN, Sezione di Trieste, Italy}
\author{Roberto Percacci}
\email{percacci@sissa.it}
\affiliation{SISSA, via Beirut 4, I-34014 Trieste, Italy, and INFN, Sezione di Trieste, Italy}
\author{Christoph Rahmede}
\email{rahmede@sissa.it}
\affiliation{SISSA, via Beirut 4, I-34014 Trieste, Italy, and INFN, Sezione di Trieste, Italy}
\begin{abstract}
We discuss the existence and properties of a nontrivial fixed point in $f(R)$--gravity, 
where $f$ is a polynomial of order up to six.
Within this seven--parameter class of theories, the fixed point has three ultraviolet--attractive 
and four ultraviolet--repulsive directions; this brings further support to the hypothesis that gravity
is nonperturbatively renormalizabile.
\end{abstract} 
\maketitle

The methods of effective Quantum Field Theory (QFT) provide an 
accurate description of low energy gravity, including calculable,
though currently unmeasurable, quantum corrections \cite{Burgess}.
On the other hand perturbative QFT fails when one tries to remove the
ultraviolet regulator \cite{Sagnotti}.
This may signal a failure of QFT to properly describe the microscopic
features of spacetime, but it is also possibile that the problems are
only related to the use of perturbation theory 
and that QFT will work when more appropriate tools are used.
Loop Quantum Gravity \cite{Rovelli} is a nonperturbative
approach based on canonical quantization methods.
Regge calculus \cite{Williams} and dynamical triangulations \cite{Ambjorn}
provide discrete nonperturbative approximations.
Here we pursue a different approach that is based on more
conventional, continuum, covariant quantum field theory methods,
and is known as ``asymptotic safety'' \cite{Weinberg}.
Loosely speaking,
a QFT is said to be asymptotically safe if there exists a finite dimensional
space of action functionals (called the ultraviolet critical surface)
which in the continuum limit are attracted towards a Fixed Point (FP) 
of the Renormalization Group (RG) flow.
For example, a free theory has vanishing beta functions, so it is a FP called the Gau\ss ian FP.
Perturbation theory describes a neighbourhood of this point.
In a perturbatively renormalizable and asymptotically free QFT such as QCD, 
the UV critical surface is parameterized by the couplings that 
have positive or zero mass dimension.
Such couplings are called ``renormalizable'' or ``relevant''.
Asymptotic safety is a generalization of this behaviour
outside the perturbative domain.

In the last ten years, evidence for the asymptotic safety of gravity has come mainly 
from the use of the Exact RG Equation (ERGE), 
describing the dependence of a coarse--grained effective action functional 
$\Gamma_k(\Phi)$ on a momentum scale $k$ \cite{Reuter,Souma,Lauscher,Perini,LauscherR2,PercacciN,Codello, Niedermaier}.
For a review see \cite{NiedermaierReuter}.
This equation has the general form 
\begin{equation}\label{ERGE}
\partial_t\Gamma_k=
\frac{1}{2}{\rm STr}\left(\frac{\delta^2\Gamma_k}{\delta\Phi\delta\Phi}+R_k\right)^{-1}\partial_t R_k
\end{equation}
where $t=\log(k/k_0)$,
$\Phi$ are all the fields present in the theory, STr is a generalized functional trace
including a minus sign for fermionic variables and a factor 2 for complex variables,
and $R_k$ is a cutoff that suppresses the contribution to the trace of fluctuations
with momenta below $k$ \cite{Wetterich}.

In the spirit of effective QFT, one assumes that $\Gamma_k[\Phi]$
has the general form
\begin{equation}
\Gamma_k[\Phi]=\sum_{i} g_i(k)\calo_i[\Phi]
\end{equation}
where $\calo_i[\Phi]$ are operators
constructed with the fields and their derivatives that have the required symmetries
and $g_i$ are running couplings of dimension $d_i$. Then
\begin{equation}
\label{gammat}
\partial_t \Gamma_k[\Phi]=\sum_{i} \beta_i(k)\calo_i[\Phi]
\end{equation}
where $\beta_i=\partial_t g_i$.
In general the functional (\ref{ERGE}) will contain infinitely many terms and infinitely many couplings;
the easiest way of extracting nonperturbative information from the ERGE
is to retain only a finite number of terms, introduce them in (\ref{ERGE}), 
evaluate the trace and read off the beta functions $\beta_i$.
Because $\partial_t R_k$ in (1) goes rapidly to zero for momenta greater than $k$,
it is not necessary to use an ultraviolet regulator in this calculation;
the beta functions are automatically finite.
We denote $\tilde g_i=g_i k^{-d_i}$ the couplings measured in units of $k$, and
\begin{equation}
\tilde\beta_i=\partial_t \tilde g_i=-d_i\tilde g_i+\beta_i k^{-d_i}\ ,
\end{equation}
where the first term comes from the classical, canonical dimension.
A FP is defined by the condition $\tilde\beta_i=0$.

The Einstein--Hilbert truncation consists 
in retaining only terms with up to two derivatives,
namely the two--parameter Lagrangian density
$g_0+g_1 R= Z(2\Lambda-R)$, 
where $Z=\frac{1}{16\pi G}$, $\Lambda$ is the cosmological constant
and $G$ is Newton's constant.
The existence of a FP in this truncation was shown in \cite{Souma}
and its properties were extensively studied in \cite{Lauscher}.
The stability of the results under changes of gauge and cutoff function
was interpreted as evidence that this FP is not an artifact of the truncation.

It can be shown in general that if a gravitational FP exists, 
Newton's constant will have an anomalous dimension equal to two \cite{Perini}, 
in such a way that the propagator of the theory has effectively
a $p^{-4}$ dependence on momentum in the UV limit.
This suggests that terms quadratic in curvature must play an important role.
The first calculation including such terms was based on the three--parameter truncation
$g_0+g_1 R+g_2 R^2$ \cite{LauscherR2}.
It was found that the FP is only slightly shifted relative to the Einstein--Hilbert 
truncation, and that it is UV--attractive in all three directions.
Subsequently, the analysis has been extended to include also the Ricci squared and
Riemann squared terms, but technical complexities 
have limited the analysis to a one--loop approximation \cite{Codello}.
In accordance with earlier perturbative calculations \cite{Julve,Buchbinder}, 
the dimensionless couplings
in the four--derivative sector were found to be asymptotically free,
while at the FP $\tilde\Lambda_*=0.221$ and $\tilde G_*=1.389$.
The nontrivial FP is attractive in all five parameters.
The results of \cite{LauscherR2} suggest that
in a more accurate treatment the FP--value 
of the four--derivative couplings would be shifted to a finite and nonzero value. 

In addition to the existence of a FP, nonperturbative
renormalizability requires that the UV critical surface,
defined as the locus of points whose trajectories are attracted towards the FP
when $t\to\infty$, has finite dimensionality.
If this condition is met, the requirement of being attracted
to the FP, which guarantees a sensible UV behaviour,
fixes all couplings up to a finite number of free parameters
that have to be determined by experiment.
This ensures that the theory will be predictive.

The attractivity properties of a FP are determined by the signs of the 
critical exponents $\vartheta_i$, defined to be minus the eigenvalues of the 
linearized flow matrix
\begin{equation}
M_{ij}=\frac{\partial \tilde\beta_i}{\partial\tilde g_j}\Bigr|_{*} .
\end{equation}
The couplings corresponding to negative eigenvalues (positive critical
exponent) are called relevant and parametrize the UV critical surface; 
they are attracted towards the FP in the UV and can have arbitrary values.
The ones that correspond to positive eigenvalues (negative
critical exponents) are called irrelevant;
they are repelled by the FP and must be set to zero.
One can show from (4) that at the Gaussian FP $\vartheta_i=d_i$,
so the relevant couplings are the ones that are
power--counting renormalizable (or marginally renormalizable).
In a local theory they are usually finite in number.

At a nontrivial FP the canonical dimensions receive loop corrections.
However, such corrections are expected to be finite,
in which case at most finitely many critical exponents could have
different sign from the canonical dimension $d_i$.
Therefore, it is generically expected that at any FP in a local theory
there will only be a finite number of relevant couplings \cite{Weinberg}.

It is clearly important to substantiate this general expectation
with explicit calculations.
Some evidence for the finite dimensionality of the UV critical surface 
comes from the leading order in the $1/N$ approximation, where the
contribution of graviton loops to the beta functions is neglected
relative to the contribution of matter fields \cite{PercacciN}. 
In this approximation the couplings with three or more powers of curvature
are irrelevant and the critical surface is five--dimensional.
In this letter we present for the first time evidence that
in pure gravity the UV critical surface is finite dimensional.

We will work with the so--called ``$f(R)$--gravity'' theories,
where the operators in (2) are restricted to be powers of the
Ricci scalar: $\calo_i=\int d^4x\sqrt{g}R^i$.
These theories have attracted much attention recently
in cosmological applications \cite{fofr}.
The quantization of such theories at one loop has been discussed in \cite{Cognola}.
Here we analyze the RG flow of this type of theories,
assuming that $f$ is a polynomial of order $n\leq 6$.
In order to apply (1) to gravity we choose a background gauge condition;
the metric is split into a background field ${\bar g}_{\mu\nu}$ and a (not necessarily small) 
quantum field $h_{\mu\nu}$: $g_{\mu\nu}={\bar g}_{\mu\nu}+h_{\mu\nu}$.
The (Euclidean) action is approximated by
\begin{equation}\label{truncation}
\Gamma_k[\Phi]= \sum_{i=0}^n g_i(k) \int d^4x\,\sqrt{g} R^i +S_{GF}+S_c\ ,
\end{equation}
where $\Phi=\{h_{\mu\nu},c_{\mu},\bar c_{\nu}\}$ and the last two terms 
correspond to the gauge fixing and the ghost sector \cite{Reuter,Dou}.
The gauge fixing will have the general form 
\begin{equation}
S_{GF}=\frac{1}{\alpha} \int d^4x\sqrt{\bar g}\,\chi_{\mu}{\bar g}^{\mu\nu}\chi_{\nu}
\end{equation}
where
$\chi_{\nu}=\nabla^{\mu}h_{\mu\nu}-\frac{1+\rho}{4}\nabla_{\nu}h^{\mu}_{\,\,\mu}$
(all covariant derivatives are with respect to the background metric) .

The ghost action contains the Fadeev--Popov term
\begin{equation}
S_c=\int d^4x\sqrt{\bar g}\,\bar{c}_{\nu}(\Delta_{gh})_{\mu}^{\nu}c^{\mu}.
\end{equation}
In the truncation (\ref{truncation}) the beta functions can be obtained 
from a calculation of the trace in the r.h.s. of (\ref{ERGE}) on a spherical 
(Euclidean de Sitter) background.

As in \cite{Dou,Lauscher,LauscherR2} we partly diagonalize the propagator by using the 
decomposition
\begin{equation}\label{decomposition}
h_{\mu\nu}=h^{TT}_{\mu\nu}+\nabla_\mu\xi_\nu+\nabla_\nu\xi_\mu+\nabla_\mu\nabla_\nu\sigma+\frac{1}{4}g_{\mu\nu}(h-\nabla^2\sigma) .
\end{equation}
The inverse propagator, including the Jacobians due to the change of variables (\ref{decomposition}), 
is given explicitly in \cite{Cognola}.
In \cite{Dou,Lauscher,LauscherR2} the Jacobians
were removed by a nonlocal redefinition of the variables $\xi$ and $\sigma$.
This works well up to $n=2$, but causes divergences for higher truncations.
For this reason in our calculation we have not performed this redefinition.
When both procedures are viable, we have checked that they lead only to very small 
numerical differences in the final results.
The Jacobians can be formally exponentiated
introducing appropriate auxiliary fields and a cutoff is introduced on these variables, too.

The cutoff operators are chosen so that the modified inverse propagator is identical to the
inverse propagator except for the replacement of $z=-\nabla^2$ by $P_k(z)=z+R_k(z)$;
we use exclusively the optimized cutoff functions $R_k(z)=(k^2-z)\theta(k^2-z)$ \cite{Litim}.
This has the advantage that knowledge of the heat kernel coefficients up to $B_8$
(which contain at most $R^4$ and which we take from \cite{Avramidi}) 
is sufficient to calculate all the beta functions.

At this point, the r.h.s. of the ERGE contains contributions from
the tranverse traceless tensor $h^{TT}$, the transverse
vector $\xi$, the two scalar components $h$ and $\sigma$,
the transverse and the longitudinal components of the ghosts, and the Jacobians.
The vector modes corresponding to the lowest eigenvalue and the scalar modes corresponding
to the two lowest eigenvalues give vanishing $h_{\mu\nu}$ in (\ref{decomposition}) and therefore
are omitted from the traces.
A considerable simplification comes from choosing the gauge $\rho=0$, $\alpha=0$: 
in this gauge the contributions from
$\xi$ and $\sigma$ cancel the contributions from the ghosts,
leaving only four terms, coming from $h^{TT}$, $h$ and the Jacobians.
Details will be given elsewhere \cite{cpr}.

The differences between our procedure and the one of \cite{LauscherR2} can then
be summarized as follows: (i) we use the optimized cutoff instead of the exponential
cutoffs; (ii) we do not redefine $\xi$ and $\sigma$; (iii) we remove one extra mode from
the ghost spectra; (iv) we use the gauge $\rho=0$, $\alpha=0$ instead of $\rho=1$, $\alpha=1/Z$.
These differences and the ensuing simplifications
allow us to calculate the r.h.s. of (\ref{ERGE}) in de Sitter space exactly:
it is a rational function of $R$ and the couplings $\tilde g_i$.
The beta functions can be extracted from
this function by taking derivatives:
\begin{equation}
\beta_i=\frac{1}{i!}\frac{1}{V}\frac{\partial^i}{\partial R^i}\partial_t \Gamma_k\Biggr|_{R=0}\ ,
\end{equation}
where $V=\int d^4v\sqrt{g}$.
This has been done using algebraic manipulation software, and
the limit $n\leq 6$ was set by the hardware 
(a standard single--processor machine).

We can now state our results.
Table I gives the position of the nontrivial FP and table II gives
the critical exponents, for truncations ranging from $n=1$
(the Einstein--Hilbert truncation) to $n=6$.
For convenience and for the sake of comparison with \cite{LauscherR2}
we also list in Table III the FP--values of the cosmological constant, Newton's
constant and their dimensionless product.

\begin{table}
\begin{center}
\begin{tabular}{l|l|l|l|l|l|l|l}	
 $n$ & $\tilde g_{0*}$ & $\tilde g_{1*}$ & $\tilde g_{2*}$ & $\tilde g_{3*}$
& $\tilde g_{4*}$ & $\tilde g_{5*}$ & $\tilde g_{6*}$\\ 
\hline
 1& 0.00523& -0.0201& & & & & \\
 2& 0.00329& -0.0127& 0.00151& & & & \\
 3& 0.00518& -0.0196& 0.00070& -0.0097& & & \\
 4& 0.00506& -0.0206& 0.00027& -0.0110& -0.00865& & \\
 5& 0.00507& -0.0205& 0.00027& -0.0097& -0.00803& -0.00335& \\
 6& 0.00505& -0.0208& 0.00014& -0.0102& -0.00957& -0.00359& 0.00246\\
\end{tabular}
\end{center}
\caption{Position of the FP 
for increasing order $n$ of the truncation.}
\label{tab:mytable1}
\end{table}
\begin{table}
\begin{center}
\begin{tabular}{l|l|l|l|l|l|l|l}	
 $n$ & $\vartheta'$ & $\vartheta''$ & $\vartheta_2$ & $\vartheta_3$
& $\vartheta_4$ & $\vartheta_5$ & $\vartheta_6$\\ 
\hline
 1& 2.38& -2.17& & & & & \\
 2& 1.38& -2.32& 26.9& & & & \\
 3& 2.71& -2.27& 2.07& -4.23& & & \\
 4& 2.86& -2.45& 1.55& -3.91& -5.22& & \\
 5& 2.53& -2.69& 1.78& -4.36& -3.76+4.88 i& -3.76+4.88 i& \\
 6& 2.41& -2.42& 1.50& -4.11& -4.42+5.98 i& -4.42+5.98 i& -8.58\\
\end{tabular}
\end{center}
\caption{Critical exponents  
for increasing order $n$ of the truncation.
The first two critical exponents are a complex conjugate pair of the form
$\vartheta'\pm\vartheta'' i$.}
\label{tab:mytable2}
\end{table}
\begin{table}
\begin{center}
\begin{tabular}{l|l|l|l}	
& $\tilde\Lambda_*$ & $\tilde G_*$ & $\tilde\Lambda_*\tilde G_*$\\
\hline

$n=1$ 				& 0.130 & 0.990 & 0.13   \\
$n=2$ 				& 0.129 & 1.56  & 0.20   \\
$n=6$ 				& 0.120 & 0.949 & 0.114  \\
\end{tabular}
\end{center}
\caption{Fixed point values of the cosmological constant and Newton's constant 
for $n=1,2,6$.}
\label{tab:mytable3}
\end{table}

Some comments are in order.
First of all, we see that a FP with the desired properties exists for all truncations. 
When a new coupling is added, new unphysical FP's tend to appear; 
this is due to the approximation of $f$ by polynomials.
However, among the FP's it has always been possible to find one for which
the lower couplings and critical exponents have values that are close to those 
of the previous truncation. That FP is then identified as the nontrivial FP
in the new truncation.

Looking at the columns of Tables I and II we see that in general the properties 
of the FP are remarkably stable under improvement of the truncation.
In particular the projection of the flow in the $\tilde\Lambda$-$\tilde G$ plane agrees
well with the case $n=1$.
This confirms the claims made in \cite{Lauscher} about the validity of the
Einstein--Hilbert truncation.

The greatest deviations seem to occur in the row $n=2$, and in the columns $g_2$ and $\vartheta_2$.
The value of $g_{2*}$ decreases steadily with the truncation. 
The critical exponent $\vartheta_2$ appears for the first time in the truncation $n=2$
with a very large value, but it decreases quickly and seems to converge around $1.5$.
This behaviour may be related to the fact that $g_2$ is classically a marginal variable.

As a further test of stability, in the truncation $n=3$
we have considered the effect of varying the gauge parameter $\alpha$ over a wide
range of values, keeping $\rho=0$.
As $\tilde\alpha$ varies between 0 and 32, $\tilde g_{i*}$, with $i=0,1,2,3$, 
vary by about 5 \%,
$\vartheta'$ changes by 8\%, $\vartheta_2$ and $\vartheta_3$ change by about 2\%.

When comparing our results for the case $n=2$ with those of \cite{LauscherR2}, one has to 
keep in mind that they generally depend on the shape of the cutoff function.
A significant quantity with very weak dependence on the cutoff function
is the dimensionless product $\Lambda G$.
The value $0.12 \div 0.14$  given in \cite{LauscherR2} for $\Lambda G$
is very close to the value we find in all truncations except $n=2$.
Our value for $\tilde g_{2*}$ in the $n=2$ truncation has the same sign 
but is between one half and one third of their value, depending on the cutoff function.
This is another manifestation of the relatively unstable behaviour of this variable.
The value given in \cite{LauscherR2} for the critical exponent $\vartheta'$ varies in the range 
$2.2 \div 3.2$
depending on the shape of the cutoff, and is in good agreement with our results,
again with the exception of the $n=2$ truncation.
Finally, in \cite{LauscherR2} the critical exponent $\vartheta_2$ has stably
large values of the order of 25 with the compact support cutoffs, 
but varies between 28 and 8 with the exponential cutoffs.
The values at the high end agree well with our result in the $n=2$ truncation.
The shape dependence that is observed with exponential cutoffs
can be taken as a warning of the truncation--dependence of this quantity.

The most important new result of our calculation is that in all truncations
the operators from $R^3$ upwards are irrelevant. 
One can conclude that in this class of truncations
the UV critical surface is three--dimensional. Its tangent space at the FP
is spanned by the three eigenvectors corresponding to the eigenvalues with
negative real part. In the parametrization (6),
it is the three--dimensional subspace in ${\bf R}^7$ defined by the equation:
\begin{eqnarray}\label{surface}
\tilde g_3&=&0.00127 + 0.190\, \tilde g_0 + 0.607\, \tilde g_1 + 1.265\, \tilde g_2\nonumber\\
\tilde g_4&=&-0.00646 - 0.732\, \tilde g_0 - 0.0156\, \tilde g_1 + 1.880\, \tilde g_2\nonumber\\
\tilde g_5&=&-0.0155 - 1.132\, \tilde g_0 - 0.846\, \tilde g_1 + 0.276\, \tilde g_2\nonumber\\
\tilde g_6&=&-0.0137 - 0.594\, \tilde g_0 - 0.932\, \tilde g_1 - 1.283\, \tilde g_2
\end{eqnarray}
Unfortunately, we cannot yet conclude from this calculation
that the operators $\calo_i$ with $i\geq3$ would be irrelevant
if one considered a more general truncation:
the beta functions that we compute for $\tilde g_i$ are really mixtures of the
beta functions for various combinations of powers of Riemann or Ricci tensors, 
which, in de Sitter space, are all indistinguishable. 
However, there is a clear trend for the eigenvalues to grow with the power of $R$.
In fact, in the best available truncation, the real parts of the critical exponents 
differ from their classical values $d_i$ by at most 2.1, and there is no tendency
for this difference to grow for higher powers of $R$.
This is what one expects to find in an asymptotically safe theory.

With a finite dimensional critical surface, 
one can make definite predictions in quantum gravity. 
The real world must correspond to one of the trajectories that emanate from the FP
and lie in the critical surface. 
Thus, at some sufficiently large but finite value of $k$ 
one can choose arbitrarily three couplings, 
for example $\tilde g_0$, $\tilde g_1$, $\tilde g_2$
and the remaining four are then determined by (\ref{surface}).
These couplings could then be used to compute the probabilities of physical processes, 
and the relations (\ref{surface}), in principle, could be tested by experiments.
The linear approximation is valid only at very high energies, but it should be possible 
to numerically solve the flow equations and study the critical surface 
further away from the FP.

Extending the results to higher polynomial $f(R)$ truncations seems to be
only a matter of computing power. In view of the results obtained here,
we expect that a FP with three attractive directions will be maintained.

\vspace{0.3cm}
\centerline{\bf Acknowledgements}

\vspace{0.5cm}
We would like to thank M. Reuter for useful conversations and correspondence.
\goodbreak

\end{document}